\documentstyle[aaspp4,11pt,psfig]{article}
\def\kms{\mbox{km s$^{-1}$}}

\def\toa{{t}}

\def\thvec{{ \mbox{\boldmath $\theta$} }}
\def\thivec{{ \mbox{\boldmath $\theta_i$} }}
\def\avec{{ \mbox{\boldmath $a$} }}

\def\fthvec{{\mbox{$f_{\scriptsize\boldmath \thvec}$}}}    
\def\fthivec{{\mbox{$f_{\boldmath \thivec}$}}}  
\def\favec{{\mbox{$f_{\boldmath\avec}$}}}      
\def\fth{{f_{\theta}}}                          

\def\ftoa{{f_{\toa}}}

\def\toaave{{\langle t \rangle}}
\def\toamax{{\toa_{\rm max}}}

\def\xvec{{\bf x}}
\def\yvec{{\bf y}}
\def\sigavec{{ \mbox{\boldmath $\sigma_a$} }}

\def\sigd{{\sigma_{\rm d}}}   

\def\pdf{{PDF}}
\def\pdfs{{PDFs}}

\def\dsj{D_{\rm s_j}}
\def\avecj{{\avec_j}}
\def\xmax{{X_{\rm max}}}
\def\thmax{{\theta_{\rm max}}}
\def\thrms{{\theta_{\rm rms}}}
\def\mathbf{{\bf}}

\def\sigth{{\sigma_{\theta}}}

\def\veff{{V_{\rm eff}}}
\def\ahat{{\bf \hat a_{\phi}}}

\def\Qprime{{Q^{\prime}}}
\def\Dsprime{{D_s^{\prime}}}

\def\dnaint{{\prod_j \int d\avecj}}
\def\dthiint{{ \int d\thivec}}

\def\lperp{{\ell_{\perp}}}
\lefthead{Cordes \& Lazio}
\righthead{Anomalous Scattering}

\received{2000 May~24}
\revised{2000 July~6}
\accepted{2000 November~8}
\paperid{MS~52007}
\begin{document}
\title{Anomalous Radio-Wave Scattering from Interstellar Plasma Structures}
\author{J. M. Cordes}
\affil{Astronomy Department and NAIC, Cornell University; 
	cordes@spacenet.tn.cornell.edu}
\author{T.~Joseph~W.~Lazio}
\affil{Code~7213, Naval Research Laboratory, Washington, DC
	20375-5351; lazio@rsd.nrl.navy.mil}
\begin{abstract}
This paper considers scattering screens that have arbitrary spatial
variations of scattering strength transverse to the line of sight,
including screens that are spatially well confined, such as disks and
filaments.  We calculate the scattered image of a point source and the
observed pulse shape of a scattered impulse.  The consequences of
screen confinement include:
(1)~Source image shapes that are determined by the physical extent of
    the screen rather than by the shapes of much-smaller diffracting
    microirregularities.  These include image elongations and
    orientations that are frequency dependent.
(2) variation with frequency of angular broadening that is much weaker 
    than the trademark  $\nu^{-2}$ scaling law (for a cold, unmagnetized
    plasma), including frequency-independent cases; and
(3) similar departure of the pulse broadening time from the usually
    expected $\nu^{-4}$ scaling law.
We briefly discuss
applications that include scattering of pulses from the Crab pulsar by
filaments in the Crab Nebula; image asymmetries from Galactic
scattering of the sources Cyg~X-3, Sgr~A*, and NGC~6334B; and
scattering of background active galactic nuclei by intervening
galaxies.  We also address the consequences for inferences about the
shape of the wavenumber spectrum of electron density irregularities,
which depend on scaling laws for the image size and the pulse
broadening.  Future low-frequency ($<100$~MHz) array observations will
also be strongly affected by the Galactic structure of scattering
material.  Our formalism is derived in the context of radio scattering
by plasma density fluctuations. It is also applicable to optical, UV
and X-ray scattering by grains in the interstellar medium.
\end{abstract}


\section{Introduction}\label{sec:intro}

Images of scattered radio sources and distorted pulses from pulsars
provide some of the most-used observables for probing microstructure
in the electron density of interstellar gas.  Over the last decade,
interstellar scattering measurements have revealed asymmetries in the
scattered images of radio sources.  These are most often interpreted
in terms of underlying anisotropy of the very small irregularities
that diffract the radiation.  That anisotropy, in turn, most likely
reflects the orientation of magnetic fields in the \ion{H}{2} gas that
contains the microstructure.  Angular broadening of compact sources
and pulse distortions due to multi-path propagation are used to probe
the amplitude of scattering and also, through the frequency scaling,
to constrain the shape of the wavenumber spectrum for the
microstructure.  Inversion of scattering observables into information
about the microstructure almost invariably relies on the assumption
that the scattering strength is uniform in directions transverse to
the line of sight.

We reconsider the assumptions used to analyze angular and temporal
broadening, in particular the assumption of uniformity 
of the scattering medium transverse to the line
of sight.  One reason is that the interstellar medium (ISM) shows
structures on a wide variety of scales and so it is reasonable to
expect manifestations of nonuniformities, at least in some directions. 
Secondly, the physics that underlies asymmetric images is quite
different if the asymmetry occurs on scales much larger than diffractive
scales, as they would if the asymmetry is caused by the large-scale
distribution of diffracting irregularities.  
Thirdly, observations of the Crab pulsar show anomalous
scalings of pulse broadening with frequency.  These are interpreted by
some as indicating that scattering occurs within the pulsar
magnetosphere rather than in a cold plasma (J.~Eilek~1997, private
communication; Hankins \& Moffett~1998; Lyutikov \& Parikh~2000).  As
we show, anomalous scalings occur quite naturally from cold plasma
extrinsic to the pulsar if the scattering region is bounded in the
transverse direction.  Additionally, future observations at low radio
frequencies of a variety of sources -- including high redshift sources
-- are expected to reveal further anomalous scattering that most
likely will be the result of confined scattering structures.  Finally,
the scaling with frequency of angular and pulse broadening is often
used to constrain the shape of the wavenumber spectrum of scattering
irregularities (e.g., Cordes, Weisberg, \& Boriakoff~1985; Fey
et al.~1991).  Weakening of the frequency dependence by confined plasma
structures would be interpreted as a steeper wavenumber spectrum.
Thus it is important to assess the role of confined structures in the
observations of scattered radio sources.

In \S\ref{sec:summary1} we discuss previous treatments of angular and
pulse broadening.  In \S\ref{sec:theory} we derive a general formalism
for scattering that takes into account arbitrary variations of
scattering strength transverse to the line of sight.  Examples are
given in \S\ref{sec:examples}.  Applications to the Crab pulsar and
other Galactic sources are given in \S\ref{sec:galactic}.  Future
observations of extragalactic sources and at low frequencies are
considered in \S\ref{sec:future}.  In \S\ref{sec:ism} we discuss
possible implications for the interstellar medium.  Finally, in
\S\ref{sec:summary}, we summarize our results.
 
\section{Past Treatments of Angular and Pulse Broadening}
\label{sec:summary1}

It is well known that the shape of a scattered impulse, viewed through
a thin (along the line of sight), infinitely-extended (transverse to
the line of sight) screen with a circularly symmetric, Gaussian angle
distribution, is a one-sided exponential function (e.g., Rickett~1990
and references therein).  Thick scattering screens produce slower rise
times, while screens containing Kolmogorov irregularities (e.g.,
Rickett~1990; Lambert \& Rickett~1999) produce decays that are slower
than exponential.  Results along these lines have been presented by
Scheuer~(1968), Williamson (1972, 1973, 1975), Lee \& Jokipii~(1975), 
and Isaacman \& Rankin~(1977).  Williamson~(1975) has shown that the
pulse broadening function from multiple, discrete screens or from a
continuous medium consists of an $n$-fold convolution of (one-sided)
exponential functions.  Williamson's result applies to the case where
the phase structure function is square-law in form.  Media with
Kolmorogov wavenumber spectra produce different shapes, though the
differences are small compared to effects we consider in this paper.
The key, implicit assumption in Williamson's analysis (and essentially
all other published results on interstellar pulse broadening; however,
see Lyne \& Thorne~1975) is that the transverse extent of the
scattering screen is arbitrarily large and that the strength of
scattering is uniform across the screen.  If these assumptions are
relaxed, quite different results emerge.

We show that several ``anomalous'' phenomena occur when scattering
structures have finite transverse extents, including:
\begin{enumerate}
\item Angular broadening from scattering or refraction that scales
anomalously with $\nu$.  In the case of radio scattering in cold
plasmas, anomalous scaling is defined as a significant departure
from a $\nu^{-2}$ scaling, which is determined by the
microphysics of the plasma.\footnote{
Scattering from irregularities with a Kolmogorov wavenumber spectrum
shows $\nu^{-11/5}$ scaling under the circumstance of moderate
scattering (cf.\ Cordes \& Lazio~1991).  We do not consider this
scaling anomalous.  When a confined structure contains Kolmogorov
fluctuations, the scaling of angular size may be considerably
shallower than $\nu^{-11/5}$.}
Departures will always consist of a {\it weaker\/} dependence on frequency.
\item Elongated (or otherwise distorted) images of point
sources that are due to
scattering but do not scale as $\nu^{-2}$.
\item Multiple imaging by multiple, discrete screens
with image intensities influenced by dilution from scattering.
\item Temporal broadening of pulsar pulses which shows a {\it weaker}
dependence on frequency than $\nu^{-4}$, in accord with
the scaling of angular broadening.\footnote{
A Kolmogorov spectrum can show a $\nu^{-22/5}$ scaling for pulse
broadening.  As with angular broadening, we do not consider this
anomalous.}
\item Replication of pulses by multiple imaging from 
an ensemble of screens.
\end{enumerate}

The goal of this paper is to discuss the impact of confined or heavily
modulated  scattering screens on some of the basic observables,
primarily pulse broadening and angular broadening.  
Therefore we do not explicitly consider refraction from large scale features
in the ISM.   We do so for two reasons.  First, purely diffractive effects
are rich enough in variety that we need to isolate the discussion to
those effects.   Secondly, our results can also be applied to some cases where
refraction is important by considering a ``renormalized'' version of
the Kirchoff diffraction integral 
(e.g., Cordes, Pidwerbetsky \& Lovelace~1986).   Renormalization
of the large scale gradient and phase curvature caused by refraction
at a given screen location can be cast as an increased or decreased
image intensity and also as a change in 
the ellipticity of the angular distribution of scattered radiation.
In the following we use
a probability density, $\favec$, to describe the scattering angles
$\avec$ from a screen.
The renormalization approach allows some of the effects of
refraction --- image shifts and shapes and  intensity changes --- to be 
absorbed into $\favec$.  Our approach considers only geometrical path
length contributions to arrival times and excludes dispersive delays,
which are associated with the screen itself.  For some situations,
dispersive delays can be important.  In this paper, the points that we
wish to make concern the geometrical phase and a complete discussion
that includes dispersive delays would distract our discussion of these
points. Consequently, we defer to another paper a complete treatment
that includes all contributions to arrival times.

\section{Probability Densities for Angle of Arrival and
	 Time of Arrival}\label{sec:theory}

We derive the image of a scattered point source and the scattered
pulse shape of an impulse by calculating, respectively, the
probability density functions (\pdfs) for the angle of arrival (AOA)
and the time of arrival (TOA).  In the following, we calculate the
effects of scattering while using some of the language and mathematics
of ray theory.  Williamson~(1975) has shown the equivalence of wave
and ray optics for some contexts, as have Cordes \& Rickett~(1998).
In this paper, we make the simplifying assumption that the TOA is
related geometrically to the AOA.  This relationship applies when only
the geometrical path length of a given ray path contributes
significantly to the TOA.  In general, the TOA includes another term
related to the integrated refractive index (the integrated electron
density in the case of a cold plasma).  We ignore the non-geometrical
term because there are astrophysical contexts where its contribution
is negligible.

Consider a series of diffracting screens at distances
$\dsj$ from a source that is  at distance $D$ from the observer. 
Letting $\avecj$ be the (two-dimensional) scattering angle from 
the $j$th screen, the angular deviation $\thvec(s)$ of a ray 
path and its transverse offset $\xvec(s)$ from the direct
ray path  at distance $s$ from the source are
\begin{eqnarray}
\thvec(s) &=& \thivec + \sum_j \avecj U(s-\dsj) \label{eq:con1} \\
\xvec(s)  &=& s\thivec  + \sum_j  (s-\dsj) \avecj U(s-\dsj),
\label{eq:xconstraint}
\end{eqnarray} 
where $U(x)$ is the unit step function.  The first equation relates
the observed ray angle  ($\thvec$) to the   
the initial ray angle ($\thivec$) and the scattering angles ($\avecj$). 
The relation $\xvec(D) = 0$ 
stipulates that rays must reach the observer.  
We assume all angles are small 
($\vert \thivec \vert, \vert \avecj \vert, \vert \thvec \vert \ll 1$)
, though it is not difficult to
extend our results to large angles.
Note that $\avecj$ is a random variable described by a distribution of
angles that is determined by diffraction (and, as mentioned above,
can also include refraction).

Including only the geometric path-length difference,
the corresponding time delay relative to the direct ray path is
\begin{equation}
\toa = \frac{1}{2c} \int_0^D ds \vert \thvec(s)\vert^2.
\end{equation}
The overall time delay also includes  dispersive components
which, as stated above, we choose to ignore because our main points
concern the effects of truncated screens on the arrival times.  

To calculate the probability density function (\pdf) of the observed
angle of arrival, $\thvec$, and the time of arrival, $\toa$, we use
Dirac delta functions to enforce Eq.~\ref{eq:con1} and $\xvec(D) = 0$
for those rays that reach the observer.  We use conditional
probabilities to include these relations and to integrate over the
\pdfs\ for the scattering angles in each screen and over the \pdf\ of
$\thivec$.  The result is simple for an isotropic source or, less
restrictively, where the \pdf\ of $\thivec$ is constant over the
relevant range of initial ray angles, $\thivec$, as we assume.  Note
that the equation $\xvec(D) = 0$ allows us to eliminate $\thivec$ as
an independent variable.

It is standard to assume the scattering strength is invariant across a
scattering screen.  Here we specify a more general description.
Consider each screen to scatter or refract rays according to a \pdf\
$\favec_j$ whose width varies arbitrarily across the screen.
Accordingly we write each screen's \pdf\ as $\favec_j(\avecj;
\xvec(\dsj))$, where $\xvec(\dsj)$ is Eq.~\ref{eq:xconstraint}
evaluated at the location of each screen, $s= \dsj$.

Let $Q$ be an observed quantity such as the AOA or TOA and let
$\Qprime(\thivec, \avecj, \dsj, D)$ 
be its value given $\thivec$, $\avecj$ and $\dsj$ (and it is implicit
that we consider all $j$ when there are multiple screens). The
\pdf\ of $Q$ is
\begin{equation}
f_Q(Q) = \frac {\displaystyle
                        \dnaint \favec_j(\avecj; \xvec(\dsj)) 
			\dthiint \fthivec(\thivec)\,
                        \delta(\xvec(D))        
                        \delta(Q-\Qprime(\thivec, \avecj, \dsj, D))
               }
               {\displaystyle
                        \dnaint \favec_j(\avecj; \xvec(\dsj))
			\dthiint \fthivec(\thivec)\,
                        \delta(\xvec(D)),       
               }
\label{eq:fQcomplete}
\end{equation} 
where the numerator is the joint \pdf\ of Q and $\xvec(D)=0$
(that rays reach  the observer), 
while the denominator is the PDF that rays reach the observer.
Using Eq.~\ref{eq:xconstraint}, we transform $\delta(\xvec(D))$
to a delta function involving $\thivec$, perform the integral over
$\thivec$, and assume that the PDF for $\thivec$, $\fthivec$, is
constant for angles of interest.   Then $\Qprime$ becomes independent
of $\thivec$ and the PDF of $Q$ becomes 
\begin{equation}
f_Q(Q) = \frac  {\displaystyle
                        \dnaint \favec_j(\avecj; \xvec(\dsj)) 
                        \delta(Q-\Qprime(\avecj, \dsj, D))
                }
                {\displaystyle
                        \dnaint \favec_j(\avecj; \xvec(\dsj))
                }.
\label{eq:fQ}
\end{equation} 

\subsection{The General Single Screen (N=1)} \label{sec:gss}

For the simple case of a single scattering screen, 
Eq.~\ref{eq:fQ} becomes
\begin{eqnarray}
f_Q(Q) &=& \frac  {\displaystyle
                        \int d\avec \favec(\avec; -\avec\Dsprime) 
                        \delta(Q-\Qprime(\avec, D_s, D)) 
                }
                {\displaystyle
                        \int d\avec \favec(\avec; -\avec\Dsprime) 
                }, \label{eq:fQ1} 
\end{eqnarray} 
where
\begin{eqnarray}
\Dsprime &=& D_s(1-D_s/D).  
\end{eqnarray} 
For the angle of arrival,  $Q=\thvec$,  $\Qprime = \avec(D_s/D)$ and
\begin{equation}
\fthvec(\thvec) = 
                        \left(\frac{D}{D_s}\right)^2 
                	\frac   {\displaystyle
                        \favec(\frac{D}{D_s}\thvec; -\thvec(D-D_s)) 
                        }
                        {\displaystyle
                        \int d\avec \favec(\avec; -\avec\Dsprime) 
                        }.
\label{eq:faoa}
\end{equation} 
For the TOA, $Q=\toa$ and $\Qprime = D_s(1-D_s/D)\vert\avec\vert^2 /
2c$ so only the magnitude, $\vert\avec\vert$, is constrained.
Transformation of $\delta(Q-\Qprime)$ to
$\delta(\vert\avec\vert-a_{\toa})$ yields
\begin{eqnarray}
\ftoa(\toa) &=& 
        \left (\frac{c}{\Dsprime} \right ) 
        		\frac   {\displaystyle
                        \int_0^{2\pi} d\phi\,  
                             \favec(a_t\ahat; -a_t\ahat\Dsprime)
                } 
                {\displaystyle
                        \int d\avec \favec(\avec; -\avec\Dsprime) 
                }, \label{eq:ftoa}
\end{eqnarray} 
where $\ahat$ is a unit vector,
\begin{eqnarray}
   \ahat    &=& \cos\phi\, \hat \xvec + \sin\phi\, \hat\yvec, 
\end{eqnarray} 
and 
\begin{eqnarray}
   a_{\toa} &=& \left( \frac{2c\toa}{\Dsprime} \right)^{1/2}. 
\end{eqnarray} 

The flux density of a source is conserved only for an infinite
screen with homogeneous statistics because only in that case is
as much flux scattered toward the observer as is scattered away. 
We define the  flux dilution factor as the ratio of the probability
that rays reach the observer to the probability for a uniform,
infinite screen: 
\begin{eqnarray}
\eta_F = \int d\avec \favec(\avec; -\avec\Dsprime),
\end{eqnarray}
equal to the denominator of Eq.~\ref{eq:fQ1}.  For a uniform,
infinite screen, $\eta_F = 1$.  In general, $\eta_F \le 1$.

We illustrate these expressions by considering specific cases.

\subsubsection{Infinitely Wide Screen with Homogeneous Statistics}\label{sec:iws}

For an infinite screen with homogeneous statistics, the denominator of
Eq.~\ref{eq:fQ1} is unity.   Specializing to circularly symmetric $\favec$, 
we find the
normalized 1D \pdf\ for the magnitude $\theta \equiv \vert \thvec \vert$:
\begin{equation}
\fth(\theta) = 2\pi\theta \left(\frac{D}{D_s}\right )^2 \favec(D\thvec/D_s).
\label{eq:pdf1c}
\end{equation}
If $\favec$ is a Gaussian function with rms angle $\sigma_a$ in each
coordinate direction, then
\begin{eqnarray}
\fth(\theta) &\approx& \sigth^{-2}
   \theta e^{-\theta^2/2\sigth^2} \\
\sigth &\equiv& (D_s/D) \sigma_a \\
\ftoa(\toa) &=& \tau_0^{-1} e^{-\toa/\tau_0}U(t)\\
\tau_0 &=& \Dsprime \sigma_a^2 / c.
\end{eqnarray}
The 1D rms angle
\begin{equation}
\thrms \equiv \frac{1}{\sqrt{2}}\langle \vert \thvec \vert^2 \rangle^{1/2}
\end{equation}
characterizes the {\it observed} range of angles and in this
case is identical to $\sigth$, which is the scaled range
of {\it scattering} angles.  In general,
$\thrms \ne \sigth$. 

If $\sigma_a \propto \nu^{-2}$ as for a plasma, 
then the  AOA \pdf\ has a scale  $\sigma_{\theta}\propto \nu^{-2}$ 
and the TOA \pdf\ has scale  $\tau_0 \propto \nu^{-4}$.   
These scaling laws for observable
quantities rely on the assumption that the screen is infinitely
wide.

\subsubsection{Circular Screen with Finite Radius}\label{sec:sns}

Consider  a circular screen with radius $\xmax$ centered on the line of sight. 
Now the \pdf\ for $\theta$ is truncated for 
$\theta > \thmax \equiv \xmax/(D-D_s)$.
Again adopting circularly symmetric, Gaussian statistics for $\avec$, we find
\begin{eqnarray}
\fth(\theta) &=& 
        	[ \sigth^{2} (1 - e^{-\zeta}) ]^{-1}
		\theta 
                e^{-\theta^2/2\sigth^2} 
		\, U(\thmax-\theta), \\
\zeta &\equiv& \frac{1}{2} \left (\frac{\thmax}{\sigth}\right)^2,
\end{eqnarray}
where the unit step function enforces truncation of the \pdf\ for
$\theta > \thmax$.   
If the rms scattering angle is small, $\sigma_a \ll \xmax D/D_s(D-D_s)$,
the scaling of the  observed size with frequency is according to that
of $\sigma_a$.  For larger scattering angles,  the physical size of the
screen becomes important.  To see this, we calculate the 
rms angular size,
which is, for the circularly symmetric Gaussian and a frequency 
scaling $\sigma_a = {\sigma_a}_0 (\nu/ \nu_0)^{-2}$, 
\begin{equation}
\thrms = 
{\sigma_a}_0 
\left(\frac{D_s}{D}\right ) 
\left(\frac{\nu}{\nu_0}\right)^{-2}
  \left [ \frac{1 - (1 + \zeta)e^{-\zeta}}
               {1 - e^{-\zeta}} \right ]^{1/2}.
\label{eq:thetarms}
\end{equation} 
The frequency scaling is $\zeta\propto \nu^{4}$.   
At large $\nu$, $\zeta \gg 1$ and $\thrms \propto \nu^{-2}$.
At small $\nu$, $\zeta \to 0$ and $\thrms$ 
becomes frequency independent.

The TOA has a \pdf\ and mean value
\begin{eqnarray}
f_{\toa}(\toa) &=& \frac{\displaystyle e^{-\toa/\tau_0}}
                      {\tau_0 \left( 1 - e^{-\zeta} \right)}  
                  U(\toamax - \toa) \\
\langle \toa \rangle &=& \tau_0 
    \left[ 1 - \zeta \left ( \frac{\displaystyle e^{-\zeta}}
                                  {1 - e^{-\zeta}} \right) \right] \\
\toamax &=& \zeta \tau_0.
\end{eqnarray}
For a wide screen, $\zeta\to \infty$, $\toamax\to\infty$ and  
$\toaave = \tau_0$, as before.  However, a narrow screen
with $\zeta \ll 1$ gives $\toaave \sim \zeta\tau_0$.
The TOA scaling is $\langle \toa \rangle \propto \nu^{-4}$ for 
wide screens but becomes frequency independent for very narrow screens. 

The observed flux of a source viewed
through the truncated screen is attenuated  by the scattering.
The flux dilution factor is
\begin{equation}
\eta_{\rm F} = 
              1 - e^{-\zeta}.
\label{eq:dilution}
\end{equation}  
For $\zeta \ll 1$, $\eta_{\rm F} \sim \zeta$ while $\zeta \gg 1$
yields $\eta_{\rm F} = 1$.  When the angular diameter and pulse
broadening of a source are observed to have anomalous frequency
dependence, i.e., when scattering is dominated by a truncated screen,
the flux is diminished.  Inspection of Eq.~\ref{eq:thetarms} and
Eq.~\ref{eq:dilution} indicates that there can be a correlation of rms
angular size and flux density.

\subsection{A Screen with Arbitrary Variations of Scattering Strength }
\label{sec:arb}

Above, we considered screens with extreme variations of scattering
strength: discontinuous or truncated to zero.  Here we consider other
cases that may have relevance to sources that are viewed through
scattering regions with structure on scales $\sim (D-D_s)\theta$.  As
in Eq.~\ref{eq:fQ}, we specify the scattering angle $\avecj$ from the
$j$th screen by a \pdf\  that depends on location along the screen,
$\xvec(\dsj)$: $\favec_j(\avecj; \xvec(\dsj))$.  For simplicity, we
drop the $j$ subscripts and discuss scattering from a single screen.
Also, for ease of discussion, we consider $\favec$ to be a 2D Gaussian
with angular variance $\sigavec(\xvec)$ that varies across the screen.

In some circumstances, the angle-of-arrival distribution
$\fthvec$ is determined purely by the shape and width of
$\favec$, while in others it is determined by the variations
in $\sigavec$ across the screen.  If $\sigavec$ is constant across
the screen, $\fthvec$ is a scaled version of $\favec$ and the
frequency scaling of the observed angular size is identical to
that of the scattering angles.

However, if $\sigavec$ varies across the screen, the observed angular
size may reflect variations of $\sigavec$ in addition to or rather
than the width of $\favec$ itself.  Let $\ell_{\avec}$ be the
characteristic length scale on which $\sigavec$ varies across the
screen.  We compare this with the observed angular diameter projected
back to the screen, yielding a length scale $\lperp \sim
(D-D_s)\thrms$.  We consider three cases:
\begin{enumerate}
\item If $\ell_{\avec}\ll \lperp$  and
variations in $\sigavec$ are statistically homogeneous, the variations
average out.  The scaling with frequency of the observed
AOA width is identical to that of $\avec$, which is due to the
microphysics. 
Note that our examples of truncated screens in previous sections satisfy
the inequality but are not statistically homogeneous. 
\item If $\ell_{\avec}\gg \lperp$,  then
$\sigavec\approx$ constant over the part of the screen sampled and the
angle-of-arrival \pdf\  is determined by $\favec$.  The frequency scaling
is again determined by the microphysics.  Depending on how large
$\ell_{\avec}$ is, eventually time variations in the image are expected
on  a time scale $t_a \sim \ell_{\avec}/\veff$, where $\veff$ is an
effective velocity determined by the velocities of the source, medium
and observer (e.g., Cordes \& Rickett~1998). 
\item If $\ell_{\avec}  \sim \lperp$, the
AOA \pdf\  is determined by a combination of $\favec$ and the spatial
variation of $\sigavec$.  The truncated screen of
\S\ref{sec:sns} is an extreme example of this case. 
The frequency scaling is likely to depart significantly from
that of the microphysics.
\end{enumerate}

\section{Examples}\label{sec:examples}

\subsection{Illustration of Anomalous Frequency Scaling}

\begin{figure}
\begin{center}
\vspace*{-1cm}
\mbox{\psfig{file=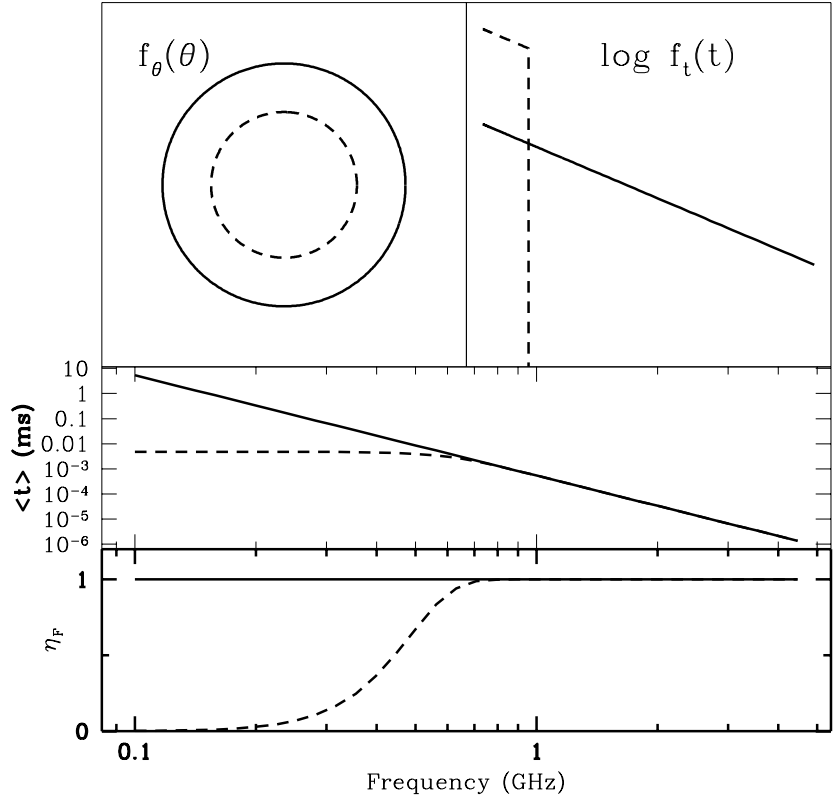,width=0.7\textwidth,silent=}}
\end{center}
\vspace*{-0.5cm}
\caption{
Scattering from an infinite screen (solid lines) and a truncated
circular screen of radius 6~mas (dashed lines).
Top left: image contours for an rms angle $\sigma_{\theta} =  10$~mas.
For the infinite screen, the contour is at 1$\sigma$ of the Gaussian
profile; for the truncated screen, the contour represents the sharp
edge of the screen.
Top right panel:  pulse broadening functions on a log-linear scale. 
Middle panel:  mean pulse broadening times plotted against frequency
for $\sigma_{\theta} = 1$~mas at~1~GHz, $D_s = 0.33$~kpc, and $D =
1$~kpc.
Bottom panel:  flux dilution factors, $\eta_F$.
}
\label{fig:inf}
\end{figure}

Figure~\ref{fig:inf} shows the ``image'' $\fthvec$, the pulse broadening function
$f_t$, and the scaling with frequency of the pulse broadening time
for two cases:
(a)~an infinite screen (bold solid lines) and
(b)~a truncated circular screen centered on the direct line of sight
   (light and dashed lines). 
The underlying scattering function ($\favec$) is a circular Gaussian
\pdf\  and the rms scattering angle scales as $\nu^{-2}$.   For the infinite
screen, the pulse broadening scales as $\nu^{-4}$, as expected.
However, the truncated screen yields truncated images and truncated
pulse broadening functions if the rms scattering angle is large enough
that rays from the screen's edges reach the observer.   Thus, 
truncation occurs at low frequencies and not at high frequencies.
For the example given, the break frequency $\sim 0.5$ GHz.
Actual break frequencies will depend on particular sizes and scattering
strengths of screens.

As a second illustration of
anomalous frequency scaling, we 
calculate the pulse broadening for a pulse scattered by a two
component screen.  The first, circular
 component with radius $X_1$
 is centered on the line of sight and scatters radiation
much more strongly than the remainder of the screen, which is of
infinite extent.   The distributions of AOA and TOA  follow from
the master equations, 
Eq.~\ref{eq:faoa},\ref{eq:ftoa}.  The rms angular size, the mean TOA,
and related quantities are:
\begin{eqnarray}
\theta_{\rm rms}&=& \sqrt 2\, \sigd_1\left(\frac{D_s}{D}\right)
\left[ \frac
        {1 + (\tau_2/\tau_1)(1+\zeta_2)e^{-\zeta_2} - (1+\zeta_1)e^{-\zeta_1}} 
        {1 + e^{-\zeta_2} - e^{-\zeta_1}} \right ]^{1/2} \label{eq:thetarms1}\\
\toaave &=& \tau_1 
\left[ \frac
        {1 + (\zeta_1+\tau_2/\tau_1)e^{-\zeta_2} - (1+\zeta_1)e^{-\zeta_1}} 
        {1 + e^{-\zeta_2} - e^{-\zeta_1}} \right ] \label{eq:tbar}\\
\tau_{1,2} &=& 
    c^{-1} \Dsprime \sigd_{1,2}^2 \\         
\zeta_1 &=& \frac{1}{2} 
        \left[ \frac{X_1}{\Dsprime\sigd_1} \right]^2 \\
\zeta_2 &=& \frac{\tau_1}{\tau_2} \zeta_1,
\end{eqnarray}
where $\sigd_{1,2}$ is the rms scattering angle produced by each screen.

\begin{figure}
\begin{center}
\vspace*{-1cm}
\mbox{\psfig{file=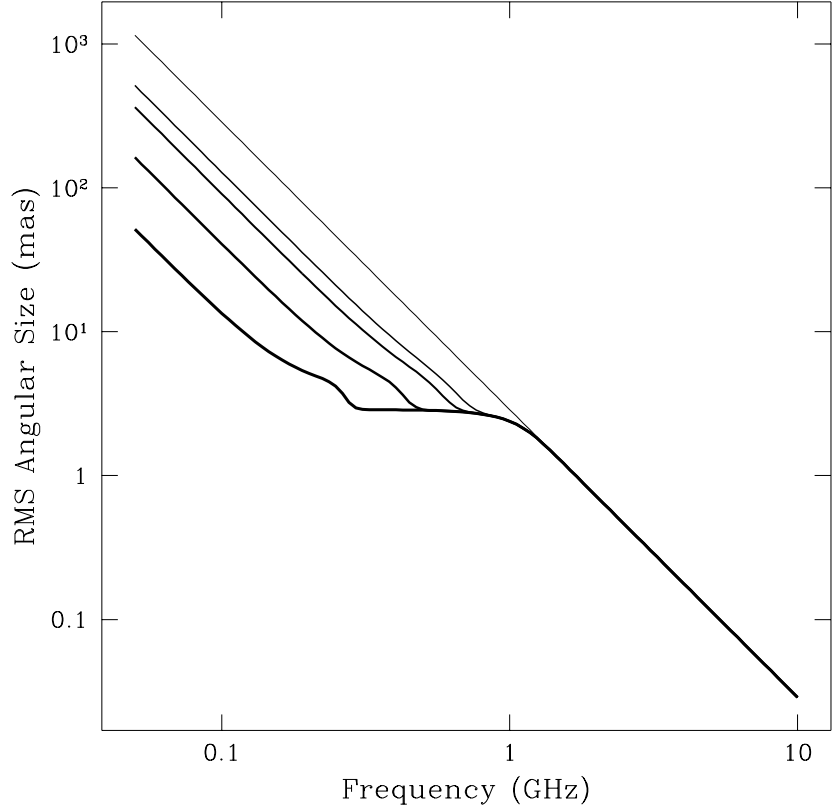,width=0.7\textwidth,silent=}}
\end{center}
\vspace*{-0.5cm}
\caption{RMS angular size plotted against frequency for
a two component scattering screen, with a strongly scattering central
part and a more weakly scattering extended part.  The curves are
evaluations of Eq.~\ref{eq:thetarms1} and correspond to different
ratios, $\sigd_1/\sigd_2= 500, 50, 10, 5, 1$ in order of thickest to
thinnest lines, where $\sigd_1$ is the rms scattering angle from the
central part and $\sigd_2$ is the value for the extended part.  In
producing this figure, we have varied $\sigd_1/sigd_2$ while keeping
$\sigd_1$ constant.  The wavy features in some of the curves are real
and reflect the different frequency dependences of terms that are
subtracted in the expression for rms angular size.}
\label{fig:example2}
\end{figure}

Figure~\ref{fig:example2} shows $\theta_{\rm rms}$ plotted against
frequency for different ratios, $\sigd_1/\sigd_2$ and assuming that
$\sigd_{1,2} \propto \nu^{-2}$.  Note that we vary $\sigd_1/\sigd_2$
while keeping $\sigd_1$ constant.  The figure demonstrates how the
stronger central component dominates the apparent source size at high
frequencies and the weaker, distributed component dominates at low
frequencies.  At intermediate frequencies, there is a plateau where
the angular size is nearly independent of frequency.

\begin{figure}
\begin{center}
\vspace*{-1cm}
\mbox{\psfig{file=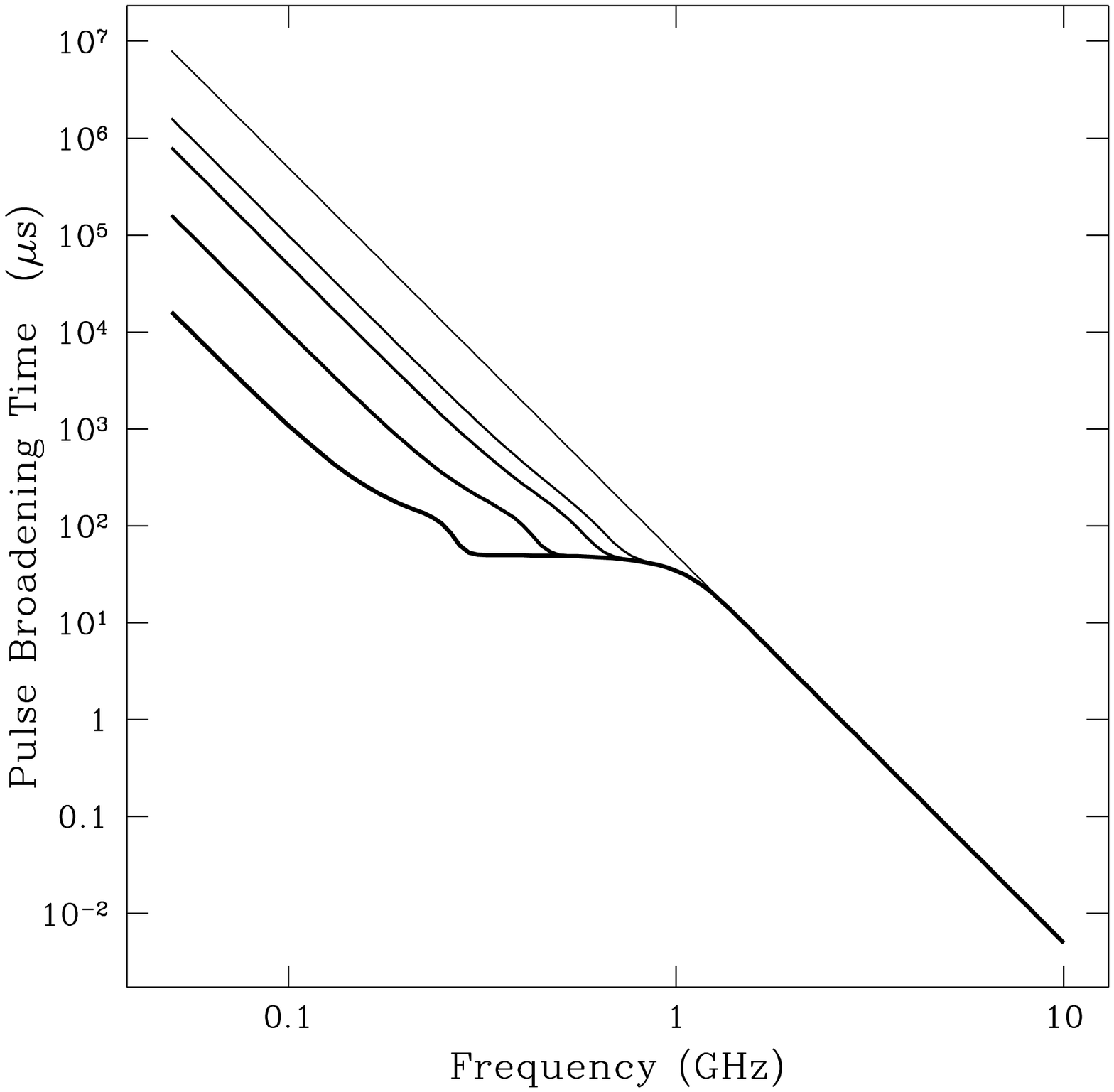,width=0.7\textwidth,silent=}}
\end{center}
\vspace*{-0.5cm}
\caption{Pulse broadening time plotted against frequency for
a two component scattering screen, with a strongly scattering
central part and a more weakly scattering extended part.
The curves are evaluations of Eq.~\ref{eq:tbar} and
 correspond to different ratios,
$\tau_1/\tau_2 = 500, 50, 10, 5, 1$ in order of thickest
to thinest lines, where $\tau_1$ is the pulse
broadening time of the central part (if it were infinite in extent)
and $\tau_2$ is the value for the extended part.
}
\label{fig:example3}
\end{figure}

Figure \ref{fig:example3} shows a similar plot, now for the mean pulse
broadening time, $\toaave$, plotted against frequency for different
ratios, $\tau_1/\tau_2$ and assuming that $\sigd_{1,2} \propto
\nu^{-2}$.  The roles of the central and distributed components are
the same as for the angular scattering shown in
Figure~\ref{fig:example2}.

\subsection{Scattering from Filaments}

\begin{figure}
\begin{center}
\mbox{\hspace{2cm}\psfig{file=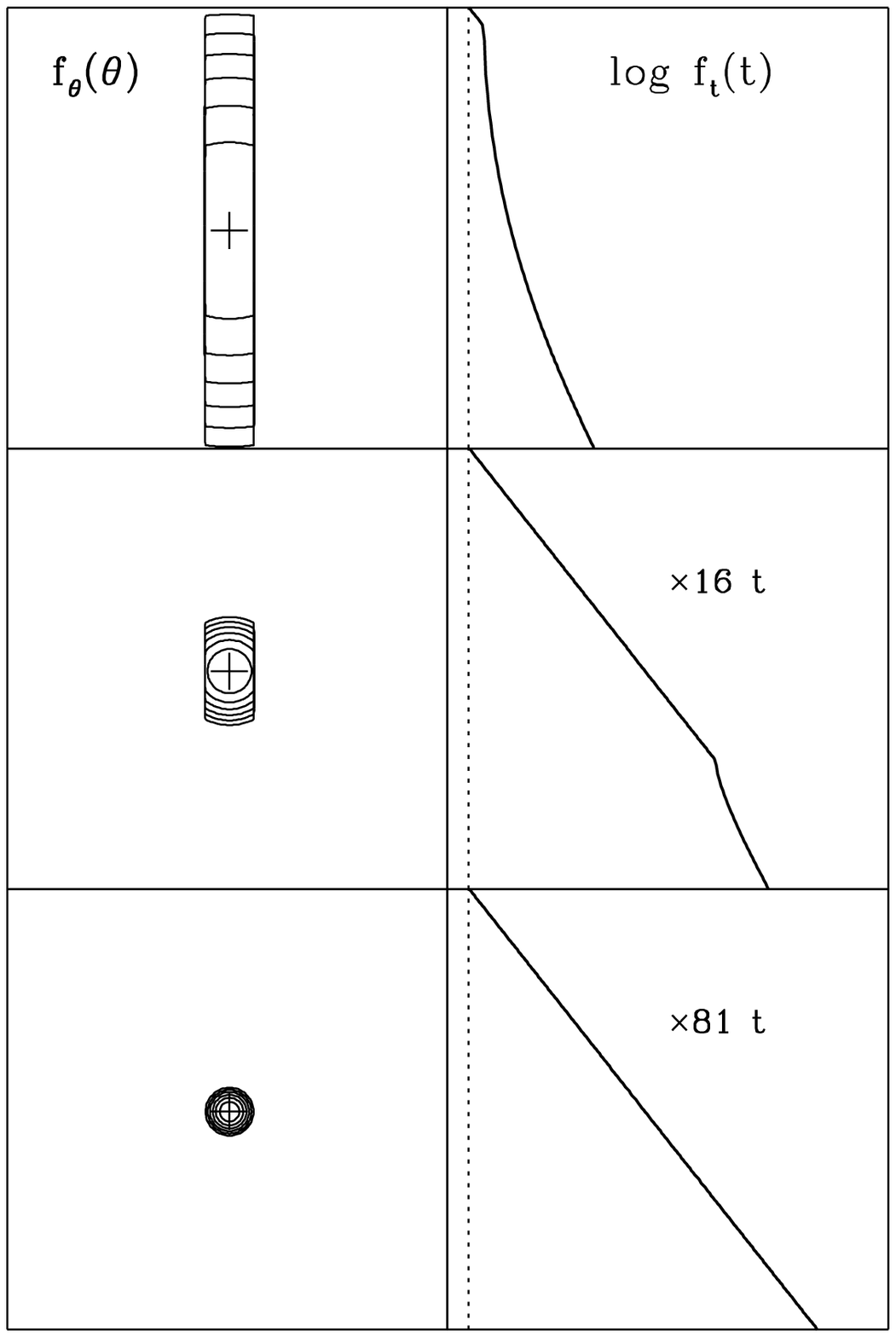,width=0.70\textwidth,silent=}}
\end{center}
\vspace*{-1.5cm}
\caption{
Images and pulse broadening functions for a vertical filament centered on the 
direct line of sight  but with different rms scattering angles.
The filament has a width of~0.28~mas, a length of~300~mas and
truncates sharply.  Equations \ref{eq:faoa},\ref{eq:ftoa} were evaluated
to obtain the plots.
Left panels: Images shown on a frame of size 2.5 $\times$ 2.5~mas.
The `+' symbol designates the direct line of sight to the source.
The rms scattering angle is constant in the filament and  
is $1\,{\rm mas}\, \nu^{-2} = 1$, 0.25, and~0.11~mas for $\nu = 1$, 2, and~3~GHz
in going from the top to the bottom frame.
The contour levels extend from the maximum down to $10^{-3}$ of the
maximum in uniformly spaced intervals of one half decade.  
Right panels:
The pulse broadening
function is shown on a log-linear scale with 1 decade 
on the vertical axis. The horizontal axis extends to 1.5~$\mu$s
in the top frame and to factors of 16 and 81 times smaller in the 
middle and lowest panels to compensate for the $\nu^{-4}$ dependence
of the pulse broadening time (if there were a continuous screen). 
The vertical dashed line indicates zero delay.
Top: Large rms scattering angle so that the image shape is dominated
     by the edges of the filament.  The pulse broadening function is
     nearly of the form $t^{-1/2}e^{-t}$.
Middle: Smaller rms scattering angle so that the image is less elongated.
        The pulse broadening function is exponential at small delays
	but shows a break point because there is an absence of large
        delays due to the truncation of the filament. 
Bottom: Rms scattering angle small enough so that the edges of the filament
        are not seen.  The pulse broadening function is exponential in
        form, $e^{-t}$.  
}
\label{fig:fil1}
\end{figure}

Figure~\ref{fig:fil1} shows scattering from a filament located along
the direct ray path for three values of rms scattering angle in the
filament.  Small rms scattering yields a circular image and an
exponential pulse broadening function.  For sufficiently large
scattering angles, the image becomes elongated and tends toward a
$t^{-1/2}e^{-t}$ pulse broadening function

\begin{figure}
\begin{center}
\mbox{\hspace{3cm}\psfig{file=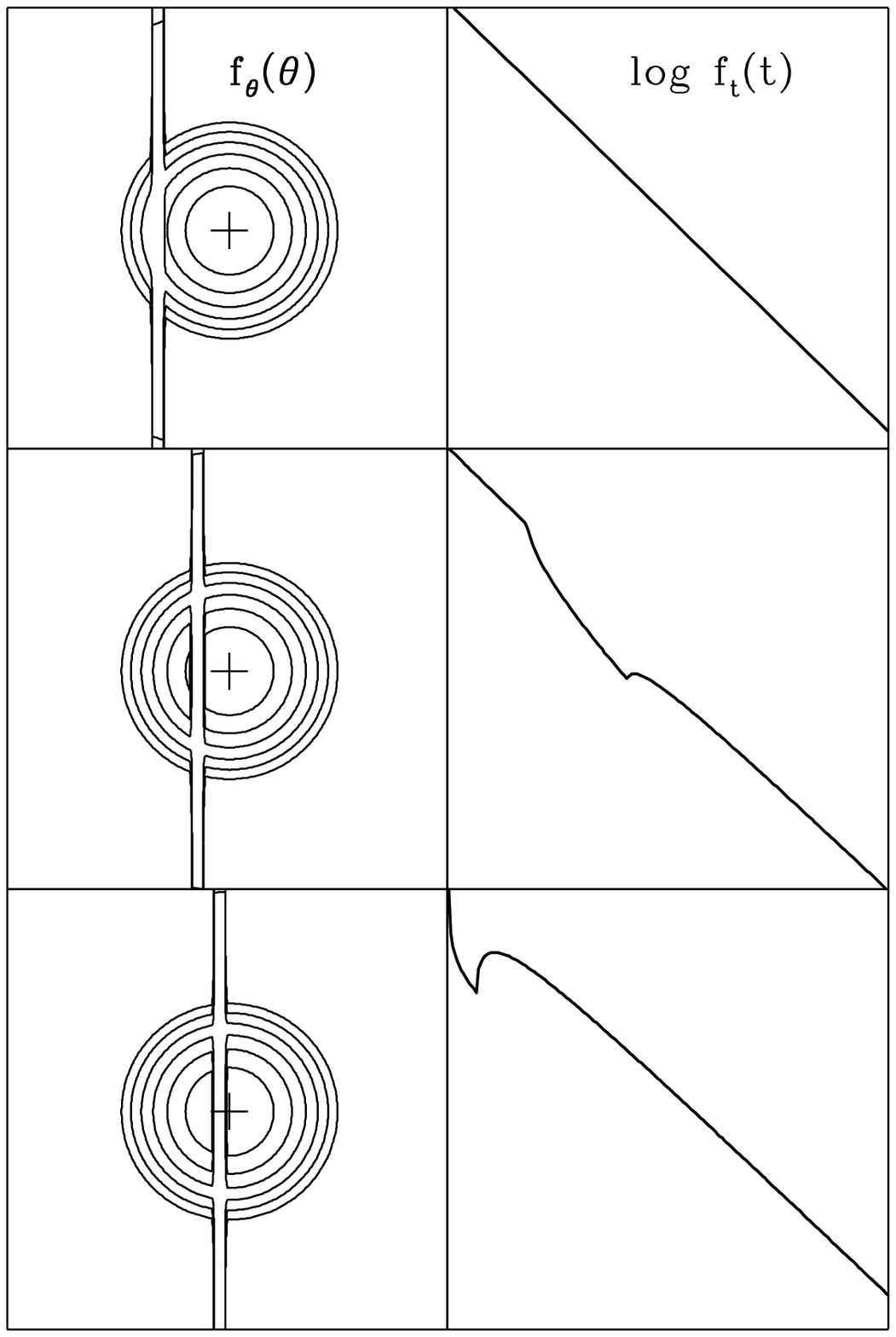,width=0.7\textwidth,silent=}}
\end{center}
\vspace*{-1.5cm}
\caption{
Images and pulse broadening functions for a filament at different
locations from the direct line of sight, evaluated using
Eq.~\ref{eq:faoa}-\ref{eq:ftoa}.  The filament is 0.075 mas wide and
300 mas long and the rms scattering angles are 4 and 0.5 mas inside and
outside the filament, respectively.   
Left panels: Images of size 2.5 $\times$ 2.5 mas 
shown with contours spaced at half-decade intervals
down to $10^{-3}$ from the peak.
The `+' symbol designates the direct line of sight to the source.
Right panels: pulse broadening functions shown on log-linear scales;
the vertical axis covers one decade and the horizontal axis
covers 0.3~$\mu$s. 
(Top)  The image is a circular Gaussian combined with a 
      subimage from the filament.  The filamentary subimage is weaker
      in amplitude because flux is diluted by the heavier scattering
      from the filament.   The pulse broadening function
      is nearly exponential in form, $\propto e^{-t}$.  
(Middle)  The filament is close enough so that its subimage now
contributes significantly to the pulse broadening function, which is
distorted by the  depletion of flux at some delays by the scattering
from the filament.
(Bottom)  The circular and filamentary subimages are sufficiently merged
so that the pulse broadening function is bimodal: a narrow component from
the portion of the filament nearest the direct line of sight and a second peak
associated with the nonfilamentary scattering.   Relative to the peak, the
pulse broadening function is now slightly broader than in the top and 
middle cases.
}
\label{fig:fil2}
\end{figure}

Figure \ref{fig:fil2} shows scattering from a filament at different
locations relative to the direct ray path but for identical rms
scattering angles in the filament.  When the filament is near enough
to the image center, the pulse broadening function is bimodal.  When
the filament is far, the pulse broadening function is dominated by the
much weaker scattering from outside the filament.  Clearly, a
superposition of filaments near the direct ray would produce a
multiplicity of pulses.

\begin{figure}
\begin{center}
\mbox{\hspace{3cm}\psfig{file=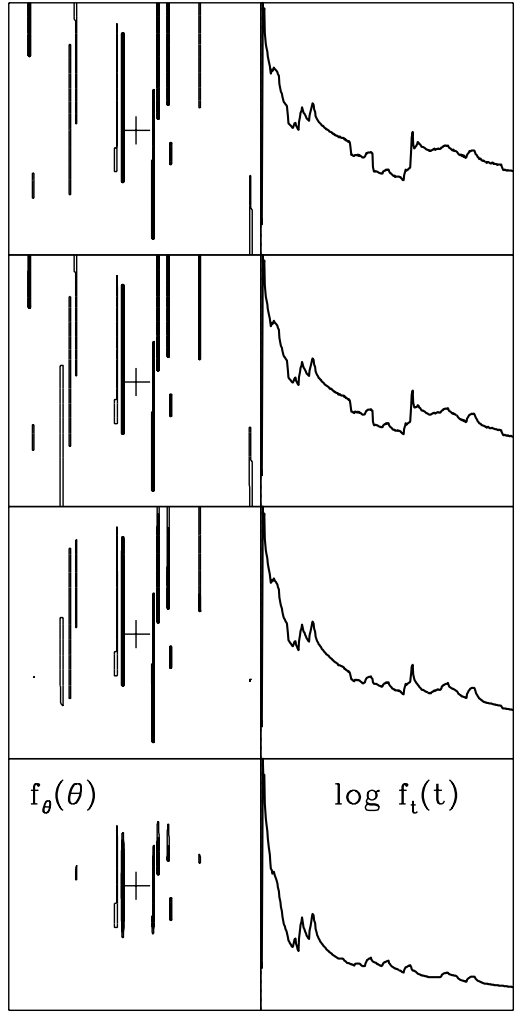,width=0.8\textwidth,silent=}}
\end{center}
\vspace*{-2cm}
\caption{
Images and pulse broadening functions for an ensemble of  filaments 
superposed with a much stronger scattering screen.
The filaments therefore represent `gaps' where the scattering is
weaker than in the screen, but still strong enough to scatter radiation
back to the observer.   
Physical filament widths and lengths are 0.03 mas and 1.5 mas, respectively,
though their apparent lengths are smaller because the flux
scattered to the observer depends on distance from the direct ray path.
At 1 GHz, the screen has an rms scattering angle of 2 mas
while the filaments have scattering angles assigned randomly
and uniformly between 0.1 and 2 mas,
Scattering angles scale as $\nu^{-2}$.
From top to bottom, $\nu = 0.1, 0.2, 0.3 $ and 0.5 GHz.   
Left panels: images are shown in 2.5 $\times$ 2.5 mas frames with
contours spaced at half-decade intervals over 1 decade from the image peaks.
Contributions from the strong screen are too small to be reflected 
in the contours.
Right panels: pulse broadening functions shown on a log linear scale
with one decade on the vertical axis. The horizontal axis covers
6.7~$\mu$s. The observed shapes are dominated by the filamentary gaps.
}
\label{fig:fils}
\end{figure}

Figure~\ref{fig:fils} shows scattering from an ensemble of filaments
at different frequencies.  As can be seen, the pulse broadening
function shows multiple peaks that align at different frequencies.
The number of filaments that are `lit up' by the scattering decreases
in going to higher frequency.  The scattering screen consists of a
very strong, extended component which has embedded ``gaps'' where the
scattering is weaker but still strong enough to scatter radiation
toward the observer.  The extended component scatters radiation to
such wide angles that it produces negligible contributions to the
pulse broadening function and to the image.  Thus the strongest
contributions to measured quantities come from the filamentary gaps.
We will explore this result further in a separate paper to discuss
giant pulses from the Crab pulsar.

\section{Application to Galactic Scattering}\label{sec:galactic}

Galactic sources show a wide range of scattering levels, indicative of
the concentration of intense scattering into a Population~I type
Galactic distribution (\cite{cwb85}; \cite{tc93}).  Here we discuss
particular objects whose scattering may be interpreted in the context
of this paper's formalism.  Our discussion is brief.  We defer to
separate articles any detailed treatment on particular sources.

\subsection{The Crab Pulsar}

The Crab pulsar shows enhanced pulse broadening from nebular material
that has been recognized since shortly after the pulsar's discovery
in 1968 (e.g., Vandenburg~1976 and references therein).  
The nebular contribution is highly episodic, with
dramatic increases of the pulse broadening time by a factor of~100
(Lyne \& Thorne~1975; Isaacman \& Rankin~1977).  Recently, multiple
images have been inferred from the presence of echoes of the pulse
shape (\cite{g_sl00}; \cite{dcb00}; Backer, Wong, \& Valanju~2000).

Giant pulses from the Crab pulsar show additional evidence for nebular
contributions to the scattering that are probably from discrete
filaments.  At relatively high frequencies (1.4 to~5~GHz), giant
pulses show multiple pulse components that tend to have exponential
scattering tails with time constants that often differ, even within
the same spin period of the pulsar (\cite{hm98}; \cite{s+99};
\cite{han00}).  Also, the widths of the pulse components appear to
scale less strongly with frequency than $\nu^{-4}$.  These
characteristics suggest consistency with the overall picture developed
in this paper.  A detailed analysis of the Crab pulsar's pulses is
deferred to another paper.

\subsection{NGC~6334B}\label{sec:ngc6334b}

The largest angular broadening measured is for the extragalactic
source NGC~6334B viewed through the \ion{H}{2} complex NGC~6334
(Trotter, Moran, \& Rodr{\'\i}guez~1998 and references therein),
$\theta_d \approx 3\arcsec$ at~1.4~GHz.  The image's position angle
rotates in going from low to high frequency, and the image axial ratio
may increase from~1.2 at~1.4 and~5~GHz to~1.5 at~15~GHz.  Trotter et
al.~(1998) interpret this variation as signifying an outer scale for
the wavenumber spectrum of \emph{anisotropically-scattering} density
irregularities, $\ell_{\rm out} \lesssim 10^{16}$~cm.  This proposed
outer scale is comparable to the thickness of the \ion{H}{2} shell in
this region (Rodr{\'\i}guez, Canto, \& Moran~1988; Kahn \&
Breitschwerdt~1990).  An alternate possibility is that the density
irregularities scatter isotropically, but that the anisotropic images
reflect density irregularities that are confined to the \ion{H}{2}
shell.

If the latter were the case, we would expect that the axial ratio
would increase as a function of decreasing frequency, opposite to what
is observed.  At lower frequencies, the size of the minor axis of the
scattering diameter would be constrained by the width of the
\ion{H}{2} shell while the size of the major axis would be essentially
unconstrained, unless the scattering is so intense that the size of
the major axis is also limited by the scale of the \ion{H}{2} shell.
For NGC~6334B, the axial ratio appears to be constant with frequency
or increasing with increasing frequency.  The frequency behavior of
the axial ratio indicates that any relevant length scales in the
\ion{H}{2} shell must be smaller than $10^{14}$~cm (the smallest
length scale probed by the highest frequency observations) or larger
than $10^{17}$~cm (the largest length scale probed by the lowest
frequency observations).

\subsection{Cygnus~X-3}\label{sec:cygx3}

The compact source Cyg X-3 is heavily scattered (0\farcs5 at~1~GHz)
and has an anisotropic image (axial ratio $\approx 1.2$ and increasing
with increasing frequency) whose position angle varies with frequency
(Wilkinson, Narayan, \& Spencer~1994; Molnar et al.~1995).  The
variation of position angle has been interpreted by Wilkinson et
al.~(1994) as due to a changing orientation of (anisotropic)
diffracting irregularities on a length scale of order $\theta(D - D_s)
\sim 0.01$~pc.  In their picture, the image asymmetry is due to
anisotropic diffracting irregularities, and the orientation change
with frequency is attributed to the spatial variation of those
irregularities.

An alternative explanation is that the diffracting irregularities are
isotropic and that the image anisotropy reflects spatial variations of
the strength of the diffracting irregularities on the scale $\theta(D
- D_s)$.  Molnar et al.~(1995) have proposed that the \ion{H}{2}
region DR~11 is responsible for the bulk of the scattering along this
line of sight.  A key difficulty with this explanation is that the
observed axial ratio increases with increasing frequency.  This is
inconsistent with the notion of scattering from a single filament but
may be consistent with scattering from a group of filaments which are
individually smaller than $10^{16}$~cm.

\subsection{Sgr~A* \& Galactic-Center OH/IR Masers}\label{sec:gc}

Galactic center sources show large scattering diameters ($\sim
1\arcsec$ at~1~GHz) and significant image asymmetries that vary across
the roughly 30\arcmin\ size of the scattering region (Lazio \&
Cordes~1998).  The major axis of the image of Sgr~A* shows a
$\nu^{-2}$ frequency dependence from~1.4 to~22~GHz (Yusef-Zadeh et
al.~1994; Lo et al.~1998), and the image itself displays no change in
its major axis, axial ratio, or orientation on time scales as long as
a decade (Lo et al.~1998; Marcaide et al.~1999).

Deviations from a $\nu^{-2}$ dependence for the major axis, minor
axis, or both are claimed at a variety of frequencies from as low as
43~GHz (Lo et al.~1998) to as high as 215~GHz (Krichbaum et al.~1998).
These deviations from the $\nu^{-2}$ dependence observed at lower
frequencies are commonly interpreted as an effect of the intrinsic
source diameter becoming important at the various frequencies.

The deviations could also be symptomatic of the size scale of the
scattering medium.  If this is the case, then the frequencies at which
the breaks occur correspond to spatial scales 
$\ell \sim 5$~AU (for a break occurring at~43~GHz) to as small as 0.17~AU
(at~215~GHz).  These spatial scales are comparable to the outer scale
inferred by Lazio \& Cordes~(1998), on the basis of a comparison of
scattering strength and thermal free-free emission.  However, the
velocities in the Galactic center ($\sim 50$~\kms\ suggest
that variations in the orientation of the image of Sgr~A* should be
seen on timescales of order $0.1\,{\rm yr}(\ell/1\,{\rm AU})(v/50\,{\rm
\kms})^{-1}$.  As mentioned above, these are not seen.  This implies
either that all striations in the medium are oriented in the same direction
or that the intrinsic source size is in fact important at high 
frequencies.  

The lack of variations in the image orientation indicates that there
are not likely to be striations or other structure in the scattering
screen on scales smaller than about~1~\hbox{AU}.  Variations do occur,
though, on much larger scales, of order 15\arcmin\ corresponding to
spatial scales of roughly 25~pc, the separations between Sgr~A*, the
various OH masers, and other scattered sources (Lazio et al.~1999).
Effects from the spatially-limited scattering described in this paper
are unlikely to be seen with the current census of Galactic center
sources.  The narrowband nature of OH masers means that the frequency 
dependence of their scattering diameters cannot be measured, and
Sgr~A* itself is obscured below~1~GHz due to free-free absorption by
\hbox{Sgr~A}.  Detection of additional radio transients (e.g., Zhao et
al.~1992) or radio pulsars (Cordes \& Lazio~1997) may allow such
effects to be detected at frequencies below~1~GHz.

\subsection{Extreme Scattering Events}\label{sec:ese}

``Extreme scattering events'' (ESE's) are events identified in the
light curves of several AGN's (Fiedler et al.~1987; 1994) and two
pulsars (Cognard et al.~1993; Maitia, Lestrade, \& Cognard~1999).  They
are roughly consistent with {\it refractive\/} defocusing and caustic
formation from discrete, small scale plasma structures (Fiedler et
al.~1987; Romani et al.~1987; Clegg et al.~1998).  However, alternative
explanations invoke the outer ionized regions of predominantly
neutral, primordial HI clouds (Walker \& Wardle~1998) or distributed
fluctuations much like those that account for the diffractive
scintillations of pulsars (Fiedler et al.~1994).  The fundamental
difference between these models is the implied gas pressure.  As
discussed further below, a discrete structure is necessarily
overpressured compared to the general ISM, so any such structures must
exist either transiently or in regions of small volume filling factor
that can support such pressure.

If ESEs result from discrete ionized structures, then effects
described in this paper should be present in the scattered image and
pulse shape (for pulsars).  This notion is little explored because few
ESEs have been identified and, given that most are seen from AGN's,
intrinsic source size effects can also diminish the appearance of
diffraction effects.  VLBI observations of the source 1741$-$038
undergoing an ESE have shown no indication of a truncated image (Lazio
et al.~2000). However, those observations were at relatively high
frequencies ($\ge 1.7$~GHz) and had limited dynamic range.  Pulse
timing observations of PSR~B1937$+$21 (Cognard et al.~1997) show no
change in the pulse shape, though, again these observations were
obtained at~1.4~GHz.  Future observations of a source undergoing an
ESE at lower frequencies (e.g., 0.33, 0.41, or~0.61~GHz) would place
much more stringent constraints on the notion that ESEs arise from
discrete ionized structures.

\section{Future Observations}\label{sec:future}

\subsection{Application to Extragalactic Scattering}\label{sec:extragal}

Scattering from extragalactic plasma can arise from the distributed
intergalactic medium (IGM), most of which is expected to be ionized,
from intervening Ly$\alpha$ clouds, and from intervening galaxies.
Of greatest relevance to this paper are the last two cases and, of
these, intervening galaxies are likely to be the more important
because of their greater column densities.  A face-on galaxy like the
Milky way will scatter radiation from a background source into an
apparent size of at least $\theta_d \sim 1\,(\nu/0.33\,{\rm
GHz})^{-2.2}$ mas (Cordes \& Lazio~1991; \cite{tc93}).  Scattering by
\ion{H}{2} regions yields even larger angles, so some background
sources, albeit at low-probability alignments, will display images
that reflect the sizes of \ion{H}{2} regions and, in some instances,
spiral arms that contain them.

Scattering from an edge-on galaxy will be about $10^3$--$10^4$ times
larger, or 1\arcsec--10\arcsec\ at~0.3~GHz.\footnote{
This large increase in the scattering diameter occurs because of the
presence of enhanced scattering regions, such as those described in
\S\ref{sec:galactic}.  In the Milky Way Galaxy, the approximate mean
free path between enhanced scattering regions is 8~kpc (\cite{tc93}).
A line of sight through the disk of a galaxy like the Milky Way is
quite likely to encounter one or more enhanced scattering regions
leading to the large increase cited.
}  The lateral scale is $\theta_d D \sim 15$--$150 \times
D_{3000}$~kpc for $D = 3000 D_{3000}$~Mpc.  Thus, near edge-on
galaxies will produce scattered images that, in part, display the
shapes of the galaxies.  At even lower frequencies, scattering
diameters from Ly$\alpha$ clouds and galaxies with significantly
smaller scattering strength will produce similar effects.  Imaging
radio observations at $\sim 0.1$~GHz will thus probe intergalactic
structures.

We defer to another paper a thorough discussion of intergalactic
scattering, taking into account cosmological expansion and evolution. 
Scattering may be able to probe the intergalactic medium at redshifts
near the reionization epoch. 

\subsection{Low-Frequency Galactic Observations}\label{sec:lowfreq}

The (nominally) strong frequency dependence of interstellar scattering
observables suggest that the anomalous scattering described here will
most likely occur at low frequencies.  High-resolution, low-frequency
instruments such as the Giant Metrewave Radio Telescope (GMRT,
\cite{a95}) and the proposed Low Frequency Array (LOFAR,
\cite{klecph00}) and the low-frequency Square Kilometer Array (SKA,
\cite{b00}) have or will have sub-arcminute resolution at frequencies
below~150~MHz.  Consequently, they may detect anomalous scattering
along lines of sight less heavily scattered than those described in
\S\ref{sec:galactic}.  Here we consider relevant lines of sight and
frequencies for which anomalous scattering is a possibility.

The relevant length scale in regions of less intense scattering may be
the outer scale of the density fluctuation spectrum~$\ell_0$.  (This
may also be the relevant length scale in intense regions, though its
value could be quite different and potentially much smaller.)  Near
the Sun (within $\approx 1$~kpc), $l_0 \sim 1$~pc (Armstrong, Rickett,
\& Spangler~1995).  It is unlikely that scattering diameters will probe
this spatial scale (i.e., $\theta_dD \sim \ell_0$) unless $\nu <
10$~MHz.  As the ionosphere becomes increasingly opaque at frequencies
$\nu < 10$~MHz, ground-based interferometric arrays will likely not be
affected by anomalous scattering in the solar neighborhood.

Toward the inner Galaxy, Galactic latitudes~$|\ell| < 50\arcdeg$,
stronger scattering than that in the solar neighborhood (but still
weaker than the intense scattering described in \S\ref{sec:galactic})
will obtain.  In this case, observations at meter wavelengths may
display anomalous scattering.

A competing effect for detecting anomalous scattering is free-free
absorption.  The density fluctuations responsible for interstellar
scattering also contribute to free-free absorption.  Sources seen
along heavily scattered lines of sight at shorter wavelengths may be
free-free absorbed at longer wavelengths.  For instance, free-free
absorption renders the Galactic center increasingly opaque for
frequencies $\nu < 1$~GHz (\cite{apeg91}).

\section{Implications for the Interstellar Medium}\label{sec:ism}

As alluded to before, the existence of compact, turbulence-containing
ionized structures is directly related to their longevity and rarity,
or filling factor, in the Galaxy.  Except for chance fluctuations from
distributed turbulence, many observed phenomena suggest the existence
of compact structures with densities that imply they are overpressured
compared to most of the \hbox{ISM}.  This is not overly surprising
because the ISM is highly dynamic and is in pressure equilibrium only
in some average sense.  It is not known which kinds of
locales (\ion{H}{2} regions, supernova shocks, etc.) provide the
largest scattering strengths.  We suggest, simply, that the observable
effects described in this paper might be used to better probe the
physical sizes of regions that produce the largest levels of
scattering.

Striations in interstellar gas densities on  sub-parsec scales are
most likely associated with magnetic fields.  On diffraction scales
$\lesssim 10^{11}$~cm, a compelling idea is that turbulence is
essentially two dimensional and that irregularities are elongated
along the field lines (Higdon~1984, 1986; Goldreich \& Sridhar~1995).
Larger scale filaments, such as those seen near the Galactic center
perpendicular to the plane of the Galaxy (\cite{y-zm87}), are also
thought to be along the local field direction.  If anisotropically
diffracting irregularities are contained in screens that are
themselves elongated in the same direction, it may be difficult to
separate (and thus identify) the two possible contributions to image
elongation.

The frequency dependence of anomalous scattering may offer a means for
identifying the cause of image elongation for a particular source if
it is heavily scattered by a single (or few) filament(s) (cf.\
Figure~\ref{fig:example2} and~\S\ref{sec:galactic}).  If the image
elongation arises from anisotropic scattering by small-scale density
irregularities, increased spatial resolution of the scattering
material (e.g., by making observations at higher frequencies) may
yield increased axial ratios as less spatial averaging is done over
the small-scale irregularities (e.g., as argued by \cite{wns94} for
the image of Cyg~X-3).  This change in axial ratio is expected only if
there are spatial variations of the orientation of scattering
turbules.  Conversely, if image elongation is produced by the
boundaries of filaments, a different frequency dependence may be seen.
A more complicated frequency dependence may result if the scattering
results from a number of smaller filaments (e.g.,
Figure~\ref{fig:fils}).

The wavenumber spectrum of electron density irregularities
is often constrained by the scaling law of angular size and
pulse broadening (and its reciprocal, the scintillation bandwidth)
with frequency.  For moderate scattering, where the dominant length
scales are between the inner and outer scales, the pulse broadening time
scales as $\nu^{-x}$ with
$x = -2\beta/(\beta-2)$, where $\beta$ is the exponent
of the three dimensional, isotropic wavenumber spectrum
(CWB85; Rickett~1990).  For a Kolmogorov spectrum, $\beta = 11/3$
and $x = 4.4$.   If irregularities are isotropic, so that $\beta = 11/3$,
but the medium is confined in the transverse direction, the actual
value of $x$ is lessened.   The value of $\beta$ inferred would be
greater than $11/3$ in this instance.   A similar trend occurs when
angular broadening is used to infer $\beta$.      It is not clear
which, if any, of the published constraints on $\beta$ are affected
by the influence of scattering-region confinement.   A detailed study
of the wavenumber spectrum is deferred to another paper.

\section{Summary}\label{sec:summary}

In this paper we have shown that radio scattering observables such as
image shapes and pulse broadening functions can be strongly influenced
by structure in the scattering medium on length scales substantially
larger than those that cause the scattering.  As such, careful
multi-frequency observations can be used to constrain properties of the
interstellar medium on scales that are typically $\sim
1$--10~\hbox{AU}.  Intergalactic scattering has not been identified
but is certainly expected from intervening spiral galaxies, probably
expected for some Lyman-$\alpha$ clouds, and may occur from
distributed ionized gas.  For intergalactic scattering, relevant
length scales can be comparable to the sizes of galaxies.  A
low-frequency VLBI survey of extragalactic sources may thus probe the
level of scattering in other galaxies and in the general intergalactic
medium.  It is also expected that scattering of radiation from
gamma-ray burst afterglows will be influenced in some cases by
intervening ionized gas in the IGM as well as in the Milky Way's
\hbox{ISM}.  These issues will be explored in separate articles.

We thank B.~Rickett for helpful discussions.  This work was supported
by NSF Grant AST 9819931 to Cornell University.  Basic research in
radio astronomy at the NRL is supported by the Office of Naval
Research.


\begin{thebibliography}{}

\bibitem[Ananthakrishnan~1995]{a95} Ananthakrishnan, S.  1995, JApAS, 16, 427

\bibitem[Anantharamaiah et al.~1991]{apeg91} Anantharamaiah, K. R.,
	Pedlar, A., Ekers, R.~D., \& Goss,	W.~M.  1991, \mnras,
	249, 262

\bibitem[Armstrong et al.~1995]{ars95} Armstrong, J. W., Rickett,
	B.~J., \& Spangler, S.~R.  1995, \apj 443, 209

\bibitem[Backer~2000]{dcb00} Backer, D.~C.  2000, in Pulsar
	Astronomy---2000 and Beyond, IAU Colloquium~177, eds.\
	M.~Kramer, N.~Wex, \& R.~Wielebinski (ASP: San Francisco) p.~499

\bibitem[Backer et al.~2000]{bwv00} Backer, D.~C., Wong, T., \&
	Valanju, J.  2000, \apj, 543, in press

\bibitem[Bower \& Backer~1998]{bb98}
Bower, G. C. \& Backer, D. C. 1998, \apj, 496, L97

\bibitem[Butcher~2000]{b00} Butcher, H.~R.  2000, in Radio Telescopes, 
	Proc. of SPIE Vol.~4015, 
	ed.\ H.~R.~Butcher 

\bibitem[Clegg, Fey \& Lazio~1998]{cfl98}
    {{Clegg}, A. W. and {Fey}, A. L. and {Lazio}, T. J. W.},
    1998,
    {\apj},
    496,
    {253},


\bibitem[Cognard et al.~1993]{c+93}
Cognard, I. et al. 1993, \nat, 366, 320

\bibitem[Cordes, Weisberg, \& Boriakoff~1985 ]{cwb85}
Cordes, J.~M., Weisberg, J.~M., \& Boriakoff, V. 1985, \apj, 288, 221.
(CWB85)
                                       
\bibitem[Cordes \& Lazio~1991]{cl91}
Cordes, J.~M., \& Lazio, T.~J. 1991, \apj, 376, 123.

\bibitem[Cordes, Pidwerbetsky \& Lovelace~1986]{cpl86}
Cordes, J.~M., Pidwerbetsky, A., \& Lovelace, R.~V.~E. 1986, \apj, 310, 737.

\bibitem[Cordes \& Lazio~1997]{lc97}
Cordes, J.~M.\ \&Lazio, T.~J.~W.  1997, ApJ, 475, 557


\bibitem[Cordes \& Rickett~1998]{cl98}
	Cordes, J.~M. \& Rickett, B. J. 1998, \apj, 507, 846 

\bibitem[Fey, Spangler \& Cordes~1991]{fsc91}
    {{Fey}, A. L. and {Spangler}, S. R. and {Cordes}, J. M.},
    1991,
    {\apj},
    372,
    {132}


\bibitem[Fiedler et al.~1987]{fdj+87}
	Fiedler, R.~L., Dennison, B., Johnston, K.~J., \& Hewish, A. 1987,
	\nat, 326, 675.

\bibitem[Fiedler et al.~1994]{fdj+94}
	Fiedler, R.~L., Dennison, B., Johnston, K.~J., Waltman, E.~B., 
	\& Simon, R.~S. 1994, \apj, 430, 581.

\bibitem[Goldreich \& Sridhar~1995]{gs95}
Goldreich, P. \& Sridhar, S. 1995, \apj, 438, 763

\bibitem[Graham~Smith \& Lyne~2000]{g_sl00} Graham~Smith, F.\ \& Lyne,
	A.~G.  2000, in Pulsar Astronomy---2000 and Beyond, IAU
	Colloquium~177, eds.\ M.~Kramer, N.~Wex, \& R.~Wielebinski
	(ASP: San Francisco) p.~503

\bibitem[Hankins~2000]{han00}
Hankins, T.~H. 2000, 
in Pulsar Astronomy---2000 and Beyond, IAU
        Colloquium~177, eds.\ M.~Kramer, N.~Wex, \& R.~Wielebinski
        (ASP: San Francisco) p.~165

\bibitem[Hankins \& Moffett~1998]{hm98}
Hankins, T.~H. \& Moffett, D.~A.  1998, \baas, 192, 570.

\bibitem[Higdon~1986]{jch86}
	Higdon, J.~C. 1986, \apj, 309, 342

\bibitem[Higdon~1984]{jch84}
	Higdon, J.~C. 1984, \apj, 285, 109

\bibitem[Isaacman \& Rankin~1977]{ir77}
Isaacman, R. \& Rankin, J.~M. 1977, \apj, 214, 214.

\bibitem[Kahn \& Breitschwerdt~1990]{kb90}
Kahn, F.\ \& Breitschwerdt, D. 1990, \mnras, 242, 209 

\bibitem[Kassim et al.~2000]{klecph00} Kassim, N.~E.,  Lazio,
	T.~J.~W., Erickson, W.~C., Crane, P.~C., Perley, R.~A., \&
	Hicks, B.  2000, in Radio Telescopes, Proc. of SPIE Vol.~4015,
	ed.\ H.~R.~Butcher, p.~328


\bibitem[Krichbaum, Witzel \& Zensus]{kwz99}
Krichbaum, T. P., Witzel, A. \& Zensus, J. A.
1999, in {\it The Central Parsecs of the Galaxy}, 
ASP Conference Series 186, eds. Falcke, H. et al., 89 

\bibitem[Krichbaum et al. ]{k+98}
Krichbaum, T.~P., et al.  1998, A\&A, 335, L106

\bibitem[Lambert \& Rickett~1999]{lr99}
Lambert, H. C. \& Rickett, B. J. 1999,
\apj, 517, 299

\bibitem[Lazio \& Cordes~1998]{lc98}
Lazio, T.J.W. \& Cordes, J. M. 1998, \apj, 505, 715 

\bibitem[Lazio et al.~1999]{l+99}
Lazio, T.~J.~W., Anantharamaiah, K.~R., Goss, W.~M., Kassim, N.~E., \&
	Cordes, J.~M.  1999, ApJ, 515, 196

\bibitem[Lazio et al. 2000]{l+00}
Lazio, T.~J.~W., et al.  2000, ApJ, 534, 706

\bibitem[Lo et al..~1998]{l+98}
Lo, K.~Y., Shen, Z.-Q., Zhao, J.-H., \& Ho, P.~T.~P.  1998, ApJ, 508,
	L61

\bibitem[Lundgren et al.~1995]{lcu+95}
Lundgren, S.~C., Cordes, J.~M., Ulmer, M., Matz, S.~M, Lomatch, S., Foster, R.~S., \& Hankins, T.~H. 1995, \apj, 453, 433.

\bibitem[Lyne \& Thorne~1975]{lt75}
Lyne, A.~G., \& Thorne, D.~J. 1975, \mnras, 172, 97.

\bibitem[Lyutikov \& Parikh 2000]{lp00}
	Lyutikov, M.\ \& Parkih, A. 2000, in 
	Pulsar Astronomy---2000 and Beyond, ASP Conference 
                    Series, Vol. 202, eds.\ M.~Kramer, N.~Wex, \&
	R.~Wielebinski (San Francisco: ASP) p.~393 

\bibitem[Maitia, Lestrade \& Cognard~1999]{mlc99}
Maitia, M., Lestrade, J.-F., \& Cognard, I.  1999, ApJ, submitted

\bibitem[Marcaide et al.~1999]{m+99}
Marcaide, J.~M., Alberdi, A., Lara, L., Perez-Torres, M.~A., \& Diamond,
	P.~J. 1999, A\&A, 343, 801

\bibitem[Molnar et al.~1995]{mmr95}
Molnar, L.~A., Mutel, R.~L., Reid, M.~J., \& Johnston, K.~J 1995, \apj,
438, 708.

\bibitem[Rickett~1990]{ric90}
Rickett, B.~J. 1990, ARA\&A, 28, 561.

\bibitem[Rodr{\'\i}guez, Canto, \& Moran~1988]{rcm88}
Rodr{\'\i}guez, L.~F., Canto, J., \& Moran, J.~M.  1988, \apj, 333, 801

\bibitem[Scheuer~1968]{sch68}
	Scheuer,P.~A.~G. 1968, \nat, 218, 920

\bibitem[Sallmen et al.~1999]{s+99}
Sallmen, S., Backer, D.~C., Hankins, T.~H. \& Lundgren, S. 1999,
\apj, 517, 460

\bibitem[Taylor \& Cordes~1993]{tc93}
Taylor, J.~H. \& Cordes, J.~M.~1993, \apj, 411, 674 (TC93) 

\bibitem[Trotter, Moran \& Rodriquez~1998]{tmr98}
	Trotter, A.S., Moran, J. M. \& Rodriguez, L. F. 1998, \apj,
	493, 666

\bibitem[Vandenberg~1976]{van76}
Vandenberg, N.~R. 1976, \apj, 209, 578

\bibitem[Walker \& Wardle~1998]{ww98}
Walker, M. \& Wardle, M. 1998, \apj, 498, 125

\bibitem[Wilkinson et al.~1994]{wns94}
Wilkinson, P.~N., Narayan, R., \& Spencer, R.~E.~1994, \mnras, 269, 67

\bibitem[Williamson 1972]{wil72}
	Williamson, I.~P. 1972, \mnras, 157, 55

\bibitem[Williamson~1973]{wil73}
Williamson, I.~P. 1973, \mnras, 163, 345

\bibitem[Williamson~1975]{wil75}
Williamson, I.~P. 1975, {Proc. R. Soc. Lond. A.}, 342, 131

\bibitem[Yusef-Zadeh \& Morris~1987]{y-zm87} Yusef-Zadeh, F.\ \&
	Morris, M.  1987, ApJ, 320, 545

\bibitem[Zhao et al.~1992]{z+92}
	Zhao, J., et al.  1992, Science, 255, 1538
\end{thebibliography}
\end{document}